\documentclass[footinbib,reprint,aps,prx,twocolumn]{revtex4-2}
\usepackage{amsmath}
\usepackage{amsfonts}
\usepackage{braket}
\usepackage[dvipsnames]{xcolor}
\usepackage{graphicx}
\DeclareGraphicsExtensions{.pdf, .png}
\graphicspath{ {./images/} }
\usepackage{hyperref}
\usepackage{ulem}

\begin{document}

\title{Second-order topological insulator under strong magnetic field: Landau levels, Zeeman effect, and magnetotransport}

\author{B.~A.\ Levitan}
\email{{\it levitanb} at {\it physics} dot {\it mcgill} dot  {\it ca}}
\author{T.\ Pereg-Barnea}
\affiliation{Department of Physics and the Centre for the Physics of Materials, McGill University, 3600 rue University, Montr\'{e}al, Quebec H3A 2T8, Canada}

\begin{abstract}
We study a three-dimensional chiral second order topological insulator (SOTI) subject to a magnetic field.  Via its gauge field, the applied magnetic field influences the electronic motion on the lattice, and via the Zeeman effect, the field influences the electronic spin. We compare two approaches to the problem: an effective surface theory, and a full lattice calculation.  The surface theory predicts a massive Dirac spectrum on each of the gapped surfaces, giving rise to Landau levels once the surfaces are pierced by magnetic flux. The surface theory qualitatively agrees with our lattice calculations, accurately predicting the surface gap as well as the spin and orbital components of the states at the edges of the surface Dirac bands. In the context of the lattice theory, we calculate the spectrum with and without magnetic field and find a deviation from the surface theory when a gauge field is applied. The energy of the lowest-lying Landau level is found closer to zero than is predicted by the surface theory, which leads to an observable magnetotransport signature: inside the surface gap, there exist different energy regions where either one or two chiral hinge modes propagate in either direction, quantizing the differential conductance to either one or two conductance quanta.
\end{abstract}
\maketitle
\section{Introduction}

\begin{figure}[t]
	\includegraphics[width=0.35\textwidth]{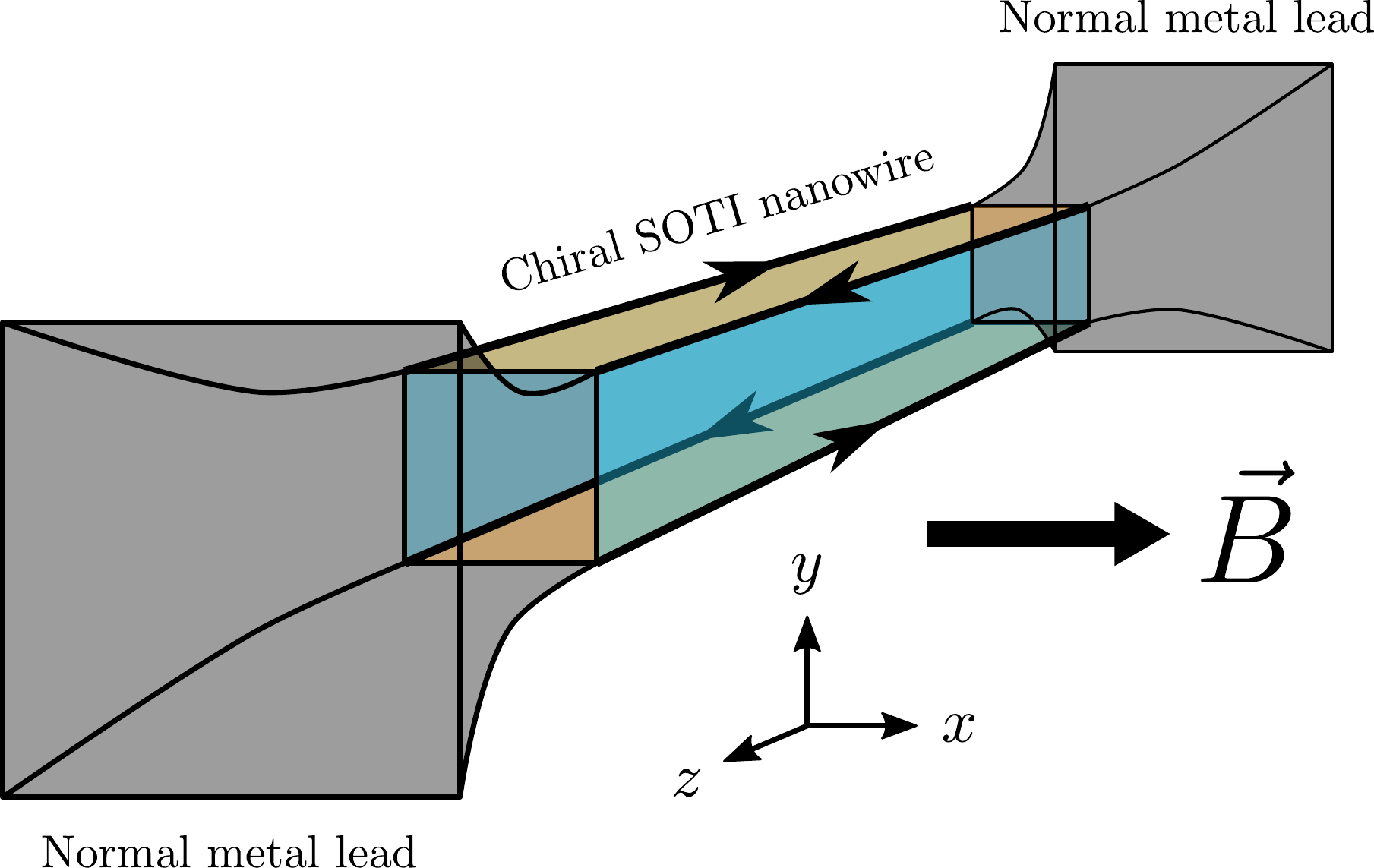}
	\caption{\label{fig:cartoon} Schematic of the system: a rectangular nanowire of chiral second-order topological insulator, subject to an external magnetic
		field $\vec{B} \parallel e_x$. To model a transport experiment, simple leads are attached to both ends of the nanowire.
		Electrons on sites at the ends of the wire  ($z = 0$ and $z = L-1$) leak into the leads at a rate $\gamma$.}
\end{figure}

The physics on the surface of a topological insulator (TI) is connected to that of its interior via the bulk-boundary correspondence \cite{KaneMele:Z2SpinHall, FuKaneMele:3DTI, FuKane:InversionTI,Hatsugai:ChernEdge,Hasan_Moore:2011Review}. In the most familiar cases, the bulk-boundary correspondence dictates that a topologically nontrivial gapped bulk is necessarily accompanied by gap-crossing surface states, protected by a combination of topology and symmetry. These surface states are typically ``anomalous" \cite{FuKaneMele:3DTI,Hasan:Colloquium}, in the sense that they cannot emerge from any lattice theory of the same dimensionality as the surface; they can only arise at the boundary of a higher-dimensional system. 

Nevertheless, if one is willing to do away with the lattice, low-energy effective continuum theories may remain useful in describing the surface of a TI. A classic example is that of a three-dimensional strong TI \cite{FuKaneMele:3DTI, FuKane:InversionTI} (3DTI), which hosts electronic surface states with gapless, Dirac-like dispersion. Each surface hosts an odd number of Dirac cones, and while no two-dimensional lattice model can produce such a spectrum without breaking time-reversal symmetry \cite{NielsenNinomiya1, NielsenNinomiya2, Marchand:2012Lattice}, the two-dimensional massless Dirac Hamiltonian $H_D (\vec{k}) = \vec{k} \cdot \vec{\sigma}$ is a useful theory for describing the physics on any one surface \cite{Zhang:TIDiracCone, Lee:TISurface, Sitte:2012TI}.

Two-dimensional surface models \cite{Schindler:HOTI, Queiroz:SplittingHinge} also provide useful insights into the three-dimensional second-order TIs (SOTIs) \cite{Benalcazar:MultipoleInsulators, Benalcazar:Pumping, Langbehn:2017Reflection, Song:2017Rotation, Geier:2018OrderTwo, Khalaf:2018Inversion, Schindler:HOTI, Khalaf:2018Symmetry, Schindler:2018Bismuth, Slager:2015Impurities}. In the case of a $C_4^z T$-symmetric chiral SOTI \cite{Schindler:HOTI}, the two-dimensional gapped surface states and the one-dimensional metallic hinge states can be thought of as a variation on the surface states of a 3DTI. While each surface of a 3DTI hosts a (gapless) Dirac cone, in the case of a SOTI, the surface Dirac states display a mass gap, with the sign of the gap alternating from one surface to its neighbour. Chiral Jackiw-Rebbi bound states then arise at the intersections between neighbouring surfaces, where the gap changes sign: these states are the hinge modes of the SOTI.

A similar two-dimensional continuum theory has been used to study the effect of an external Zeeman field on the hinge modes of a helical three-dimensional SOTI \cite{Queiroz:SplittingHinge}, coupled to the spin degree of freedom of the electrons. Interestingly, such a treatment reveals that the helical hinge modes can be split into spatially-separated pairs of counter-propagating chiral modes, and a quantum anomalous Hall effect is predicted in the region contained by the chiral modes. Especially in light of the recent experimental demonstration of Josephson interferometry between SOTI hinge channels in proximity to a superconductor \cite{Schindler:2018Bismuth}, it is natural to inquire whether other magnetically induced effects may be observed in a SOTI in the presence of magnetic flux. One might wonder whether quantum (non-anomalous) Hall physics, with its associated Landau levels, may also be possible on the surface of a SOTI. We are therefore motivated to consider a full three-dimensional model, and to include a magnetic field which couples to the orbital motion of the electrons as well as their spin. For simplicity, we focus on the case of the chiral SOTI, which possesses an odd number of chiral modes on each hinge.

\section{Model}
We begin with the tight-binding lattice model of a three-dimensional chiral SOTI introduced by Schindler \textit{et al.}\ \cite{Schindler:HOTI}, and introduce 
an applied external magnetic field in two ways. The field couples to the electronic spin via a Zeeman term $\vec{b} \cdot \vec{\sigma}$, 
and to the electronic orbital motion via the usual Peierls substitution \cite{Hofstadter:Butterfly}, which is equivalent to shifting the momentum by the gauge field (as in the standard continuum minimal coupling). The Hamiltonian is given by
\begin{multline}	\label{eq:hamiltonian}
	H = \sum_{\vec{r}} \Bigg \lbrace c^{\dagger}_{\vec{r}} \left[ M \sigma_0 \tau_z  + ( \vec{b} \cdot \vec{\sigma} ) \tau_0 \right] c_{\vec{r}}	\\
		+ \sum_{j = x,y,z} \left( c^{\dagger}_{\vec{r} + \hat{e}_j} 
			\left[ \frac{e^{i \theta_{\vec{r}}^{j}}}{2} \left(
			t \sigma_0 \tau_z + i \Delta_1 \sigma_j \tau_x
			\right) \right] c_{\vec{r}} + \mathrm{h.c.} \right)		\\
		+ \left( c^{\dagger}_{\vec{r} + \hat{e}_x} \left[ \frac{e^{i \theta_{\vec{r}}^x}}{2}  \Delta_2 \sigma_0 \tau_y \right] c_{\vec{r}} \right.	\\
		- \left. c^{\dagger}_{\vec{r} + \hat{e}_y} \left[ \frac{e^{i \theta_{\vec{r}}^y}}{2} \Delta_2 \sigma_0 \tau_y \right] c_{\vec{r}} 
			+ \mathrm{h.c.} \right)
		\Bigg \rbrace.
\end{multline}
where  we have set $\hbar = c = | \text{electron charge} | = \text{lattice constant} = 1$, and where
$\sigma_j$ and $\tau_j$ ($j \in \lbrace x, y, z \rbrace$) are the Pauli matrices for the spin and on-site orbital degrees of
freedom respectively. $c_{\vec{r}}$ is a vector of fermion annihilation operators for the $(2$ spin$) \times (2$ orbital$) = 4$ states at site $\vec{r}$. 
The Peierls phase for nearest-neighbour hopping from site $\vec{r}$ to site $\vec{r} + \hat{e}_{j}$ is $\theta_{\vec{r}}^j = - \int_{\vec{r}}^{\vec{r} + \hat{e}_j} \vec{A} \cdot \vec{\mathrm{d} \ell}$. Note that we will choose $M, \Delta_1, \Delta_2 > 0$ and $t < 0$ in order to locate the surface Dirac cones at $k = 0$.

For a finite sample with open boundary conditions in all directions, with $C_4^z T$-respecting surface terminations, and without any applied magnetic field ($\vec{A} = \vec{b} = 0$), the Hamiltonian given by Eqn.~\eqref{eq:hamiltonian} is gapped in the bulk, as well as on the four two-dimensional surfaces parallel to the $z$-axis. In the higher-order topological phase (corresponding to $1 < | M / t | < 3$ and $\Delta_2 \ne 0$), the hinges between such two-dimensional surfaces host one-dimensional chiral metallic modes. For this particular model, the two-dimensional surfaces orthogonal to the $z$-axis are gapless; this is not necessarily the case for all SOTIs. We will ultimately consider transport along the $z$-direction, as shown schematically in Fig.~\ref{fig:cartoon}, with leads coupling to the entire surface at each end of a nanowire; the metallic surfaces at each end will thus have no impact on transport.

Turning on the applied magnetic field, we focus on the case where the field points along the $x$-direction (so $\vec{b} \parallel \hat{e}_x$), and choose the Landau gauge $\vec{A} = (0, 0, B y)$. This yields Peierls phases $\theta_{\vec{r}}^x = \theta_{\vec{r}}^y = 0$ and $\theta_{\vec{r}}^z = -B y$. Note that the applied field breaks the $C_4^z T$ symmetry which protects the higher-order topological phase. We will see that the gap-crossing hinge modes nonetheless survive, but are modified.

\section{Surface theory}	\label{sec:surface_theory}

By definition, the bulk of a three-dimensional SOTI is gapped. The only states inside the bulk gap are the one-dimensional (metallic) hinge modes, and possibly the two-dimensional surface modes (depending on the size of the surface gap). It is therefore productive to derive a two-dimensional theory describing the surface states. To do so, we adapt the procedure outlined in e.g.\ Ref.~\cite{Qi:RMPTI} for the case of an ordinary 3DTI, where the surface states are massless, to the SOTI case, where the surface states are massive. We consider a surface normal to the $x$-axis, and divide the bulk Hamiltonian into two pieces describing motion parallel and perpendicular to the surface respectively: $H(\vec{k}) = H_{\parallel} (\vec{k}_{\parallel}) + H_{\perp} (k_{\perp})$. $\vec{k}_{\parallel} = (0, k_y, k_z)$ is the component of the momentum parallel to the surface, while $k_{\perp} = k_x$ is the component normal to the surface. With no applied field ($\vec{A} = \vec{b} = 0$), Fourier transforming Eqn.~\eqref{eq:hamiltonian} and taking the continuum limit by expanding up to $O(k^2)$ yields
\begin{subequations}		\label{eq:perp_and_parallel_hamiltonians}
	\begin{multline}	\label{eq:parallel_hamiltonian}
		H_{\parallel} = - \frac{t}{2} (k_y^2 + k_z^2) \sigma_0 \tau_z + \Delta_1 (k_y \sigma_y + k_z \sigma_z) \tau_x	\\
			+ \frac{\Delta_2}{2} k_y^2 \sigma_0 \tau_y
	\end{multline}
	and
	\begin{equation}
		H_{\perp} = (\tilde{M} - \frac{t}{2} k_x^2) \sigma_0 \tau_z + \Delta_1 k_x \sigma_x \tau_x - \frac{\Delta_2}{2} k_x^2 \sigma_0 \tau_y,
	\end{equation}
\end{subequations}
where we have denoted $\tilde{M} = M + 3t$. Note that $H_{\parallel} (\vec{k}_{\parallel} = 0) = 0$. To find the effective surface theory at long length scales, we first find the eigenstates at $\vec{k}_{\parallel} = 0$ and then project the Hamiltonian onto those eigenstates, keeping terms to leading order in $k_x$ and $k_y$.

We imagine a sample which is semi-infinite in the $x$-direction, with chiral SOTI occupying $x < 0$ and vacuum elsewhere. To find the states localized at the surface, we make the ansatz $\ket{\psi} = e^{\kappa x} \ket{\phi}$, with $\mathrm{Re} \kappa > 0$. Replacing $k_x \rightarrow -i \partial_x$, at $\vec{k}_{\parallel} = 0$, the Schr\"{o}dinger equation becomes
\begin{multline}	\label{eq:lambda_energy_eigenvalue}
	E \ket{\phi} = 	\\
	\left[ (\tilde{M} + \frac{t}{2} \kappa^2) \sigma_0 \tau_z - i \Delta_1 \kappa \sigma_x \tau_x + \frac{\Delta_2}{2} \kappa^2 \sigma_0 \tau_y \right] \ket{\phi}.
\end{multline}
The only spin operator to appear is $\sigma_x$, so $\ket{\psi}$ is a $\sigma_x$ eigenstate. Treating $\sigma_x = \pm 1$ as a c-number and taking the determinant yields a quartic equation in $\kappa$, relating $E^2$ and $\kappa^2$:
\begin{equation}	\label{eq:lambda_quartic}
	-\frac{1}{4} (t^2 + \Delta_2^2) \kappa^4 + (\Delta_1^2 - \tilde{M} t) \kappa^2 + E^2 - \tilde{M}^2 = 0,
\end{equation}
which has four solutions. Assuming that $\Delta_1^2 \ne \tilde{M} t$ and $E^2 < \tilde{M}^2$ is sufficient to ensure that two of the solutions have positive real part; call these $\kappa_1$ and $\kappa_2$. These solutions lead to states that obey the vanishing boundary condition at $x\to -\infty$. Since Eqn.~\eqref{eq:lambda_quartic} depends only on $E^2$ and not on $E$ itself, for each value of $\kappa$, there is an associated eigenvector for the positive putative energy $E$, and another eigenvector for the negative putative energy $-E$. To determine which solutions are physical, we make use of the boundary condition $\ket{\psi (x=0)} = 0$. This boundary condition requires a linear combination of states, $\ket{\psi} = \alpha_1 e^{\kappa_1 x} \ket{\phi_{\kappa_1}} + \alpha_2 e^{\kappa_2 x} \ket{\phi_{\kappa_2}}$. For such a superposition to meet the boundary condition, the vector $\ket{\phi}$ must be the same for both values of $\kappa$, up to an overall phase. Writing
\begin{equation}
	\ket{\phi} = \ket{\sigma_x} \otimes \begin{pmatrix}
		\phi_{0, \sigma_x}	\\
		\phi_{1, \sigma_x}
	\end{pmatrix},
\end{equation}
the matching condition can be written as
\begin{subequations} 	\label{eq:matching_condition}
    \begin{equation}	
    	\frac{\phi_{1, \sigma_x}}{\phi_{0, \sigma_x}} \Bigg{\rvert}_{\kappa = \kappa_1} = \frac{\phi_{1, \sigma_x}}{\phi_{0, \sigma_x}} \Bigg{\rvert}_{\kappa = \kappa_2},
    \end{equation}
    which, in light of Eqn.~\eqref{eq:lambda_energy_eigenvalue}, translates to
    \begin{equation}
    	\frac{\tilde{M} + \frac{t}{2} \kappa_1^2 - E}{\Delta_1 \kappa_1 \sigma_x + \frac{\Delta_2}{2} \kappa_1^2}
    		= \frac{\tilde{M} + \frac{t}{2} \kappa_2^2 - E}{\Delta_1 \kappa_2 \sigma_x + \frac{\Delta_2}{2} \kappa_2^2}.
    \end{equation}
\end{subequations}
Substituting explicit expressions for $\kappa_{1, 2}$, one finds that the matching condition of Eqn.~\eqref{eq:matching_condition} leads to $E_{\text{right surface}} = -\Delta_{\text{surf}} \sigma_x \equiv \frac{\tilde{M} \Delta_2}{\sqrt{t^2 + \Delta_2^2}} \sigma_x$.  We therefore have two surface states at $k_{||} = 0$ with the spin (anti)parallel to the $\hat x$-direction at (negative) positive energy.

Denoting the energy eigenstates as $\ket{\psi_{\mu}} = (\alpha_1 e^{\kappa_1 x} + \alpha_2 e^{\kappa_2 x}) \ket{\phi_{\mu}} \propto (e^{\kappa_1 x} - e^{\kappa_2 x}) \ket{\phi_{\mu}}$ ($\mu = 1, 2$), the effective long-distance theory of the surface is obtained by linearizing the Hamiltonian of Eqn.~\eqref{eq:parallel_hamiltonian} with respect to $k_y$ and $k_z$, and projecting onto the states $\ket{\psi_{\mu}}$. The states $\ket{\psi_{\mu}}$ can be written as
\begin{subequations}
	\begin{equation}
		\ket{\psi_1} \propto (e^{\kappa_1 x} - e^{\kappa_2 x}) \ket{\sigma_x = -1} \otimes
			\begin{pmatrix}
				i \chi_1	\\
				1
			\end{pmatrix}
	\end{equation}
and
	\begin{equation}
		\ket{\psi_2} \propto (e^{\kappa_1 x} - e^{\kappa_2 x}) \ket{\sigma_x = +1} \otimes
			\begin{pmatrix}
				i \chi_2	\\
				1
			\end{pmatrix},
	\end{equation}
\end{subequations}
where $\chi_1$ and $\chi_2$ are real parameters, meaning that the 2-component spinor describing the orbital degree of freedom lies in the $\tau_y$-$\tau_z$ plane. For a 3DTI, where $\Delta_2 = 0$, $\chi_{1,2} = \pm 1$, and so the orbital spinor lies along $\pm \tau_y$. When $\Delta_2 \ne 0$, $\chi_{1, 2}$ depend on all of the model parameters. These wavefunction $\ket{\psi_{1,2}}$ decay into the bulk exponentially with an inverse length scale $\bar{\kappa} = (\kappa_1 + \kappa_2 )/2 = \sqrt{\Delta_1^2 / (t^2 + \Delta_2^2)}$. In the basis $\left\lbrace \ket{\psi_1}, \ket{\psi_2} \right\rbrace$, the effective surface Hamiltonian is then
\begin{equation}	\label{eq:surface_hamiltonian}
	H_{\text{surface}} = 
	\begin{pmatrix}
		\Delta_{\text{surf}}	&	- v_0 (i k_z + k_y)	\\
		- v_0 (-i k_z + k_y)	&	-\Delta_{\text{surf}}
	\end{pmatrix}.
\end{equation}
The velocity $v_0$ is defined by
\begin{multline}
	\bra{\psi_1} H_{\parallel} \ket{\psi_2} = \frac{-i \Delta_1 (\chi_1 - \chi_2)}{\sqrt{(1 + \chi_1^2)(1 + \chi_2^2)}} (k_z - i k_y)	\\
		+ O(k^2)	\\
	\equiv - v_0 (i k_z + k_y) + O(k^2).
\end{multline}
$v_0$ can be interpreted as the effective ``speed of light" for the emergently-relativistic surface Dirac fermions.

The preceding analysis considered a surface termination at the right-hand-side of a sample, which required that $\mathrm{Re} \kappa > 0$. If instead we consider a termination at the left, i.e.\ $\mathrm{Re} \kappa < 0$, the sign of the energy (relative to $\sigma_x$) is reversed: $E_{\text{left surface}} = + \Delta_{\text{surf}} \sigma_x = - E_{\text{right surface}}$. On one surface normal to the x-axis, the positive-energy state at $\vec{k}_{\parallel} = 0$ has $\sigma_x = +1$ while the negative-energy state has $\sigma_x = -1$. On the opposite surface, the spins for positive- and negative-energy states are reversed, as are the associated orbital spinors $\propto (i \chi \, \, \, \, \, 1)^T$. 

\subsection{Surface Zeeman effect}
From the preceding analysis, it follows immediately that the addition of a Zeeman field ($b \ne 0$) modifies the size of the gaps on the left- and right-hand surfaces in opposite ways:
\begin{align}		\label{eq:surface_gap}
\begin{split}
	E_{\text{left/right surface}} &= (b \pm \Delta_{\text{surf}} ) \sigma_x	\\
		&= \left( b \pm \frac{(-\tilde{M}) \Delta_2}{t^2 + \Delta_2^2} \right) \sigma_x,
\end{split}
\end{align}
where the choice of $+$ ($-$) is for the left (right) surface.  Therefore the effect of a Zeeman field in the $x$-direction is to increase the gap on one surface while decreasing it on the opposite (parallel) surface.  As we'll see in Sec.~\ref{sec:spectrum} and in Fig.~\ref{fig:positionweighted_spectra}a, this is supported by our lattice calculation.

\subsection{Surface theory: Landau levels}
One might expect that the effects of an applied gauge field on the surface states of a SOTI could be captured by performing the minimal substitution $\vec{k} \rightarrow \vec{k} + \vec{A}$ in the effective massive Dirac Hamiltonian of Eqn.~\eqref{eq:surface_hamiltonian}. Such a substitution yields the familiar theory of a massive two-dimensional Dirac electron moving in a perpendicular magnetic field \cite{Haldane:1988, Sitte:2012TI, Asaga:2020MagneticDirac}. Consider the surface at the right-hand side of the sample. With a magnetic field in the positive $x$-direction, and corresponding gauge field $\vec{A} = (0, 0, By)$, the minimal substitution yields
\begin{equation}		\label{eq:surface_hamiltonian_gaugefield}
	H_{\text{surface}} = 
	\begin{pmatrix}
		\Delta_{\text{surf}}	&	- i v_0 \sqrt{2B} a^{\dagger}	\\
		i v_0 \sqrt{2B} a	&	-\Delta_{\text{surf}}
	\end{pmatrix},
\end{equation}
where we have introduced the lowering operator $a = \sqrt{\frac{B}{2}} \left( y + \frac{i}{B} k_y + \frac{1}{B} k_z \right)$, satisfying $\left[ a, a^{\dagger} \right] = 1$. This Hamiltonian admits Landau level solutions:
\begin{subequations}   
    \begin{equation}	\label{eq:zeroLL}
    	\ket{\text{$0^{th}$ L.L.}} = \begin{pmatrix}
    		\ket{0}	\\
    		0
    	\end{pmatrix},
    \end{equation}
    \begin{equation}
    	\ket{\text{$+ n^{th}$ L.L.}} = \begin{pmatrix}
		u_{+n, 1} \ket{n}	\\
		u_{+n, 2} \ket{n-1}
	\end{pmatrix},
    \end{equation}
    and
    \begin{equation}
    	\ket{\text{$- n^{th}$ L.L.}} = \begin{pmatrix}
		u_{-n, 1} \ket{n}	\\
		u_{-n, 2} \ket{n-1}
	\end{pmatrix}.
    \end{equation}
\end{subequations}
For $n \ge 1$, the coefficients satisfy
\begin{equation}
	\frac{u_{\pm n, 1}}{u_{\pm n, 2}} = \frac{\Delta_{\text{surf}} \pm E_n}{i v_0 \sqrt{2 B n}}
		= \frac{i v_0 \sqrt{2 B n}}{\Delta_{\text{surf}} \mp E_n},
\end{equation}
where the energies have magnitude
\begin{equation}
	E_n = \sqrt{\Delta_{\text{surf}}^2 + 2 B n v_0^2}	 \; \; \; (n \ge 0).
\end{equation}
In the case of the surface of a 3DTI, where the surface Dirac electrons are massless (i.e.\ $\Delta_{\text{surf}} = 0$) the lowest-energy (``zeroth") Landau level is at zero energy, and for $n \ge 1$, the $\pm n^{\text{th}}$ Landau level is at energy $\pm \sqrt{2 B n v_0^2}$. For the case of the surface of a SOTI, where the Dirac electrons acquire a mass (i.e.\ $\Delta_{\text{surf}} \ne 0$), the zeroth level is shifted to $E_0 = \Delta_{\text{surf}}$, while the other levels at $\pm E_{n \ge 1}$ shift symmetrically away from zero energy. We will compare these predictions to a full three-dimensional model in Sec.~\ref{sec:spectrum}.

\section{Spectrum}	\label{sec:spectrum}
We now turn our attention to the exact lattice model with the Hamiltonian of Eqn.~\eqref{eq:hamiltonian}. To visualize the spectrum of a chiral SOTI, we consider a sample which is periodic in the $z$-direction, and which is open in the $x$- and $y$-directions ($0 \le x \le L_x - 1$, $0 \le y \le L_y - 1$). Periodicity along $z$ eliminates the gapless surface states which would exist with open boundary conditions along $z$. Fourier transforming Eqn.~\eqref{eq:hamiltonian} along the $z$-direction only, the Hamiltonian block diagonalizes over $k \equiv k_z$, with blocks of dimension $4 L_x L_y \times 4 L_x L_y$:
\begin{equation}
	H =  \sum_{k} \sum_{x, y, x^{\prime}, y^{\prime}} c_{x y k}^{\dagger} \big( \mathcal{H} (k) \big)_{xy, x^{\prime} y^{\prime}} c_{x^{\prime} y^{\prime} k}.
\end{equation}
We assume a magnetic flux along the $x$ axis, with a flux per unit cell in the $yz$-plane equal to a rational multiple of $2 \pi$, $\vec{B} = (2 \pi p / q, 0, 0)$, with $p, q \in \mathbb{Z}$ and $L_y$ taken to be a multiple of $q$. The Hamiltonian admits the symmetry
\begin{equation}	\label{eq:symmetry}
	\mathcal{H} (k) = - \tau_x \sigma_z I_y \mathcal{H} (2 \pi p / q - k) I_y \sigma_z \tau_x,
\end{equation}
where the reflection $I_y$ exchanges sites $(x, y)$ and $(x, L_y - 1 - y)$. Therefore, we expect the spectrum to be symmetric around $k^{*} = \pi p / q$

\begin{figure}[htp]
	\includegraphics[width=0.45\textwidth]{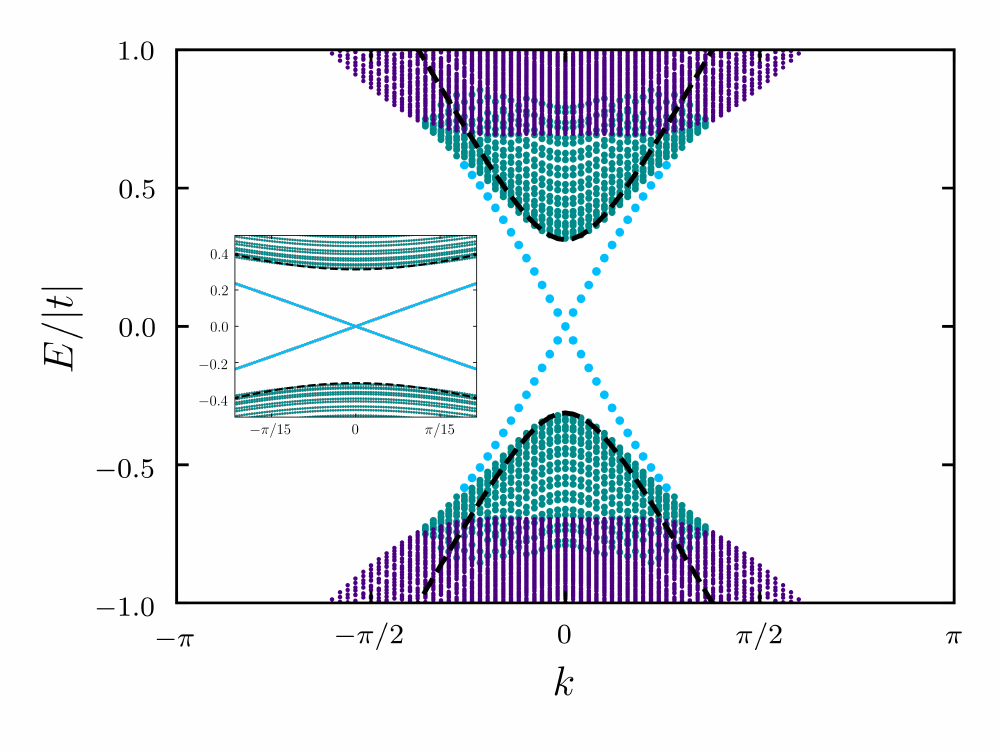}
	\caption{\label{fig:zerofield_labeled_spectrum} Energy eigenvalues of a chiral second-order topological insulating nanowire with no external field. The bulk (purple, 
		gapped), surface (turquoise, gapped), and hinge (light blue, gapless) modes are visible.
		The two visible hinge modes are each doubly-degenerate, for a total of four hinge modes (one on each hinge).
		The wire is periodic along $z$. The parameters used are $t = -1.0$, 
		$M = 2.3$, $\Delta_1 = 0.8$, $\Delta_2 = 0.5$, $\vec{b} = 0$, and $B = 0$.
		The cross-section of the nanowire is $L_x \times L_y = 30 \times 30$. 
		An effective continuum surface theory (black dashed curve) can be constructed which accurately predicts the surface gap, and the surface dispersion 
		at small $k$ (inset). The surface theory curve is obtained by diagonalizing the effective Hamiltonian of Eqn.~\eqref{eq:surface_hamiltonian} at $k_y = 0$; see 
		Sec.~\ref{sec:surface_theory}.}
\end{figure}

Fig.~\ref{fig:zerofield_labeled_spectrum} shows the spectrum of a periodic SOTI nanowire with $L_x = L_y = 30$ with no external field, obtained by numerical diagonalization. The gapless hinge modes are visible, along with the gapped surface and bulk states. The states are categorized according to their probability density in the cross-sectional plane of the wire. States with more than $50\%$ of their probability density within $3$ sites of the corners are labeled as hinge states, and plotted in light blue. States with more than $50\%$ of their probability density within $3$ sites of the surfaces, and which are not already labeled as hinge states, are labeled as surface states, plotted in turquoise. All remaining states are labeled as bulk states, and are plotted in indigo. Note that the surface gap is accurately predicted by the continuum result of Eqn.~\eqref{eq:surface_gap}, and the surface dispersion is accurately predicted for small $k$.

\begin{figure*}[htp]
	\includegraphics[width=\textwidth]{{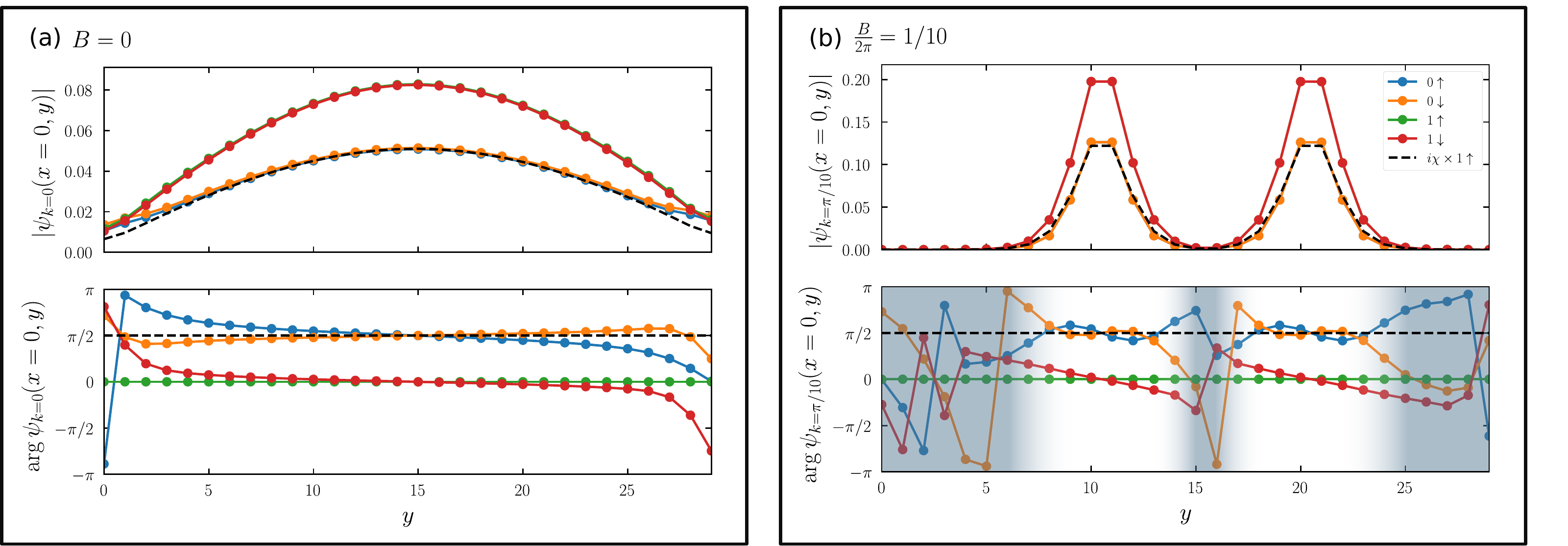}}
	\caption{\label{fig:spinor} Surface cuts at $x = 0$ through wavefunctions of SOTI surface states, in magnitude (top panels) and phase (bottom panels), for a sample 
		periodic along $z$. $0$ and $1$ ($\uparrow$ and $\downarrow$) label the orbital $\tau_z$ (spin $\sigma_z$) basis states.
		(a) When no magnetic field is applied, each surface of the SOTI hosts a massive Dirac cone. The spin and orbital state at the bottom of the 
		Dirac cone is in good agreement with the surface theory of Sec.~\ref{sec:surface_theory}, which predicts that the spin lies along 
		the $\sigma_x = +1$ direction, and that the orbital components are related by a factor of $i \chi$ (black dashed curve).
		(b) When a gauge field is applied, the surfaces pierced by flux host Landau levels. 
		The spin and orbital state of the lowest-lying Landau level matches that at the bottom of the Dirac cone from which the Landau level emerged, as expected from 
		the continuum surface theory. Note that where the Landau level wavefunction is small in magnitude, the phase becomes ill-defined (shaded regions). In both (a) and (b),
		where the green $1 \uparrow$ (blue $0 \uparrow$) component is invisible, it is obscured by red $1 \downarrow$ (orange $0 \downarrow$) component. 
		All phases are measured with respect to the $1 \uparrow$ component. The parameters used are $t = -1.0$, 
		$M = 2.3$, $\Delta_1 = 0.8$, $\Delta_2 = 0.5$, and $\vec{b} = 0$. The cross-section of the nanowire is $L_x \times L_y = 30 \times 30$.
}
\end{figure*}

Fig.~\ref{fig:spinor}a shows that the continuum surface theory also accurately predicts the spin and orbital components of the states at the edges of the massive Dirac bands. At the left hand surface of the sample (near $x = 0$), the continuum surface theory predicts the state at the bottom of the positive-energy Dirac cone to be $\ket{\phi} \propto \ket{\sigma_x = +1} \otimes (i \chi \, \, \, \, \, 1)^T$.

\begin{figure*}[htp]
	\includegraphics[width=\textwidth]{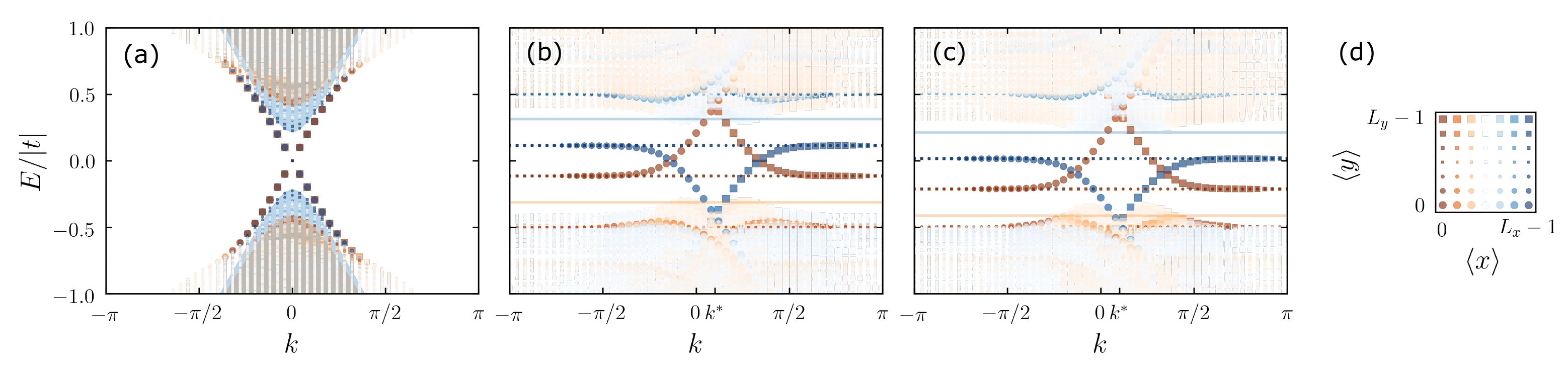}
	\caption{\label{fig:positionweighted_spectra} Energy eigenvalues of a chiral second-order topological insulating nanowire, subject to 
		(a) Zeeman field only ($\vec{b} = 0.1 e_x$, $B = 0$), 
		(b) gauge field only ($\vec{b} = 0$, $B = 2 \pi \times p / q = 2 \pi \times 1 / 10$),
		and (c) both gauge and Zeeman fields ($\vec{b} = 0.1 e_x$, $B = 2 \pi \times 1 / 10$). 
		The wire is periodic along $z$. For each state, the average position in the cross-sectional plane of the wire is represented as shown in (d), where color, shape and size represent the localization of the states. 
		Energies corresponding to a state with average $y$-coordinate $0 \le \langle y \rangle < (L-1) / 2$ are marked with circles; squares denote 
		$(L-1)/ 2  \le \langle y \rangle \le L-1$. Larger markers indicate positions farther from the centre, i.e.\ closer to $\langle y \rangle = 0$ (large circles) 
		and $\langle y \rangle = L -1$ (large squares). Colour indicates the average x-coordinate, $\langle x \rangle$. The spectra are symmetric around 
		$k^{*} = \pi \times p / q$; see Eqn.~\eqref{eq:symmetry}. The parameters used are $t = -1.0$,  $M = 2.3$, $\Delta_1 = 0.8$, and 
		$\Delta_2 = 0.5$. The cross-section of the nanowire is $L_x \times L_y = 30 \times 30$. 
		The gap-crossing hinge modes visible in (a) are doubly-degenerate, for a total of four hinge modes (one on each hinge).
		The predictions of the continuum theory for the left (right) surface at $x = 0$ ($x = L_x - 1$) are shown in pale orange 	
		(blue) solid colours, consisting of massive Dirac cones in (a), and the lowest Landau levels in (b) and (c); see Sec.~\ref{sec:surface_theory}. 
		Note that while the continuum theory accurately predicts the surface gaps
		in (a), it overestimates the energy of the lowest Landau levels in (b) and (c).}
\end{figure*}

Fig.~\ref{fig:positionweighted_spectra} shows how introducing a magnetic field modifies the spectrum of the nanowire. A Zeeman field along $x$ modifies the gaps on each of the surfaces normal to the $x$-axis in opposite ways, as expected from Eqn.~\eqref{eq:surface_gap}. An applied gauge field produces more drastic changes in the structure of the spectrum, which becomes symmetric around the nonzero $k^{*}$ (see Eqn.~\eqref{eq:symmetry}). As is the case without any applied field, each hinge hosts a mode which passes through zero energy, and these hinge modes connect gapped states on the surfaces. In $k$-space, the applied gauge field separates these modes into two pairs: the zero crossing for two modes moves to the left of the high-symmetry momentum $k^*$, while the zero crossing for the other two modes moves to the right of $k^*$. More dramatically, on the two surfaces pierced by the field ($x = 0$ and $x = L_x -1$), the applied gauge field transforms the gapped Dirac surface states into Landau levels. Applying a gauge field to the continuum surface theory of Sec.~\ref{sec:surface_theory} would predict that the energy of the lowest Landau level should be away from zero energy, with its energy equal in magnitude to the mass of the Dirac electrons from which that Landau level emerges. In fact, on both surfaces pierced by flux, while the lowest Landau level is indeed away from zero energy, its energy is reduced in magnitude relative to the Dirac mass $\Delta_{\text{surf}}$. The reduction in energy of the lowest Landau level increases with increasing magnetic flux, i.e.\ with \textit{decreasing} magnetic length. We conjecture that this reduced mass may be related to the physics of Hofstadter's butterfly \cite{Hofstadter:Butterfly}, originating in the competition between magnetic and lattice length scales, explaining the discrepancy between the continuum theory and the lattice model.

Fig.~\ref{fig:spinor}b shows that the continuum surface theory accurately predicts the spin and orbital components of the zeroth Landau level. In particular, the surface theory predicts that the spin and orbital components should match those of the state at the bottom of the Dirac cone from which the Landau level emerges; see Eqn.~\eqref{eq:zeroLL} and surrounding discussion.

The Landau levels display an additional feature of note. In the spectrum of Fig.~\ref{fig:positionweighted_spectra}b, the levels corresponding to $n > 0$ restricted to either $x = 0$ or $x = L-1$ alone are not symmetric around zero energy. Rather, the Landau level at energy $E$ has its counterpart at $-E$ on the opposite face of the wire.

\section{Magnetotransport}	\label{sec:magnetotransport}

Without external field, each hinge of a chiral SOTI nanowire hosts an odd number of current-carrying metallic modes, which cross through the gap, connecting the valence and conduction surface bands \cite{Schindler:HOTI}. In the simplest case, there is one such mode on each hinge, with propagation directions alternating as one moves around the nanowire. Therefore, in the gap, there are two modes propagating in each direction; when the chemical potentials of the leads are in the bulk and surface gaps of the SOTI, the differential conductance of such a nanowire would be expected to be $2 e^2 / h$. 

In Sec.~\ref{sec:spectrum} we showed that while a continuum surface Dirac theory successfully predicts the formation of Landau levels on a SOTI surface pierced by magnetic flux, the energies of the predicted Landau levels differ from those obtained exactly from a tight-binding model. In particular, the Landau levels on the SOTI surface exist at lower energies than the mass of the Dirac electrons which give rise to those Landau levels. As is the case without the gauge field, the hinge modes still connect the negative-energy (valence) and positive-energy (conduction) surface states. Therefore, there exists a range of energies in between the lowest Landau level and the gap of the dispersing surface states (which are localized near $y = 0$ and $y = L-1$), where only \textit{one} hinge mode exists in each propagation direction along $z$ (see Fig.~\ref{fig:positionweighted_spectra}b). We therefore expect the quantization of the differential conductance in that energy range to change, from $2 e^2 /h$ to $e^2 /h$, when the external gauge field is turned on.

To verify the preceeding intuition, we calculate the differential conductance of a nanowire in the Landauer formalism \cite{Datta:Transport}. The differential conductance from the left (taken to be $z = 0$) to the right (taken to be $z = L-1$), at a given energy, is directly proportional to the transmission probability at that energy, summed over all channels. The sum of transmission probabilities is given by
\begin{equation}	\label{eq:landauer}
	T(E) = \mathrm{Tr} \, {G^R \Gamma_L G^A \Gamma_R}.
\end{equation}
$G^{R (A)}$ is the retarded (advanced) Green function of the nanowire, including the effects of the coupling to the leads:
\begin{equation}
	G^R (E) = \left( E - H + i \eta - \Sigma^R (E) \right)^{-1},
\end{equation}
and $G^A = (G^R)^{\dagger}$. $\Gamma_{L (R)}$ describes the loss of electrons into the left (right) lead, and is directly related to the self-energy induced on the nanowire by the lead:
$\Gamma_{L (R)} = i \left( \Sigma^{R}_{L (R)} - \Sigma^{A}_{L (R)} \right)$. We use the simplest possible lead self-energies; for the left lead,
\begin{equation}
	\left( \Sigma^{R}_{L} \right)_{xyz, x^{\prime} y^{\prime} z^{\prime}} = \frac{-i \gamma}{2} \sigma_0 \tau_0 \delta_{z, 0} \delta_{z, z^{\prime}} \delta_{x, x^{\prime}} \delta_{y, y^{\prime}}.
\end{equation}
where $\gamma$ is an inverse lifetime scale. $\Sigma^{R}_{R}$ is given by an analogous expression, only replacing $\delta_{z, 0}$ by $\delta_{z, L_z - 1}$. These self-energies correspond to the assumption that the only effect of the leads on the system is that electrons at the system-lead interfaces can leak out of the system.

\begin{figure}
	\includegraphics[width=0.45\textwidth]{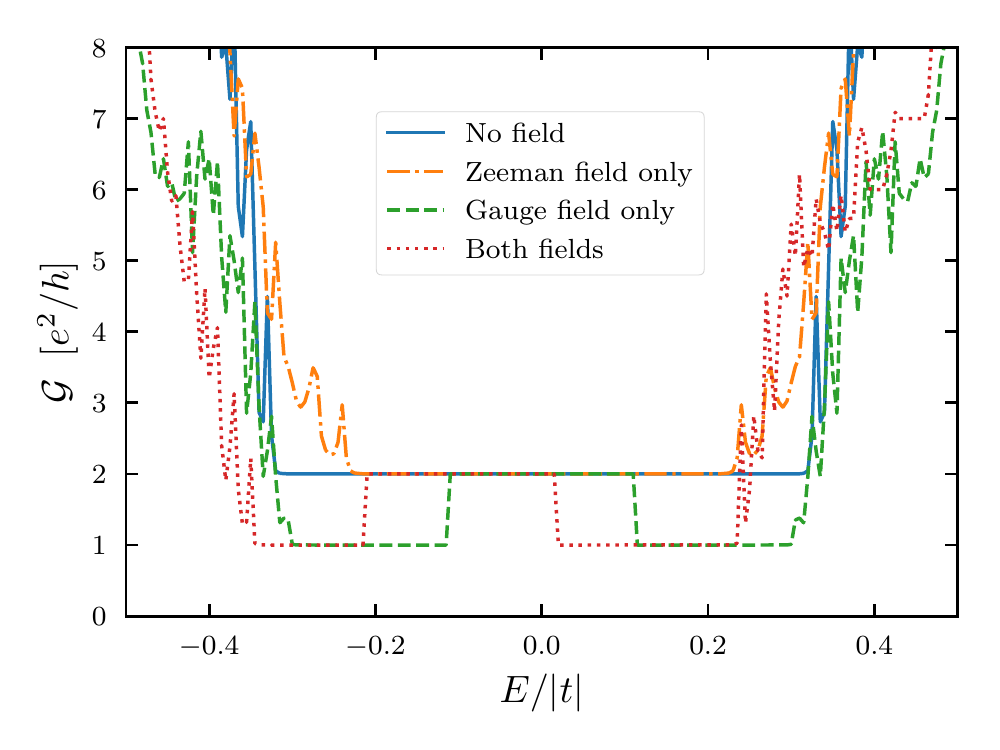}
	\caption{\label{fig:transport} Differential conductance of a chiral second-order topological insulating nanowire, with and without magnetic 
		field. Within the gap, transport is quantized to the number of available hinge modes. The sample has
		$30 \times 30 \times 30$ sites, with open boundary conditions. The parameters used are the same as for Figs.\ \ref{fig:zerofield_labeled_spectrum}, \ref{fig:spinor}, and
		\ref{fig:positionweighted_spectra}.
		}
\end{figure}

The differential conductance is obtained from \eqref{eq:landauer} via $\mathcal{G} (E) = (e^2 / h) T(E)$. Results for a finite sample (dimensions $L_x \times L_y \times L_z = 30 \times 30 \times 30$) are shown in Fig.~\ref{fig:transport}. Within the bulk and surface gaps, the conductance is quantized to the number of hinge modes available propagating in the $+z$ direction. As expected from the spectra in Fig.~\ref{fig:positionweighted_spectra}, this quantization changes from $2 e^2 / h$ when there is no external gauge field, to $e^2 / h$ when the external gauge field is applied, for energies in between the lowest Landau level and the gap of the dispersing surface states $\Delta_{\text{surf}}$. We stress that this change in quantization would not be expected if the surface electrons were simple two-dimensional Dirac electrons in a perpendicular magnetic field. In that case, the lowest Landau level energies would be the same as the gaps on the surfaces parallel to the applied field, and conductance quantization to $2 e^2 / h$ would be expected throughout the gap.

\section{Conclusion}
We have studied a three-dimensional chiral second order topological insulator in the presence of an external magnetic field.  We analyzed both the response of the electronic spin, due to the Zeeman effect, and the response of the electronic orbital motion, due to the external gauge field added as a Peierls phase in electron hopping terms.  Before solving the model numerically on a finite lattice, we have derived an effective theory for the states localized to the surfaces of the sample.  These have massive Dirac-like dispersion. The mass changes sign between neighbouring surfaces, giving rise to the hinge modes. We compared the spectrum and wavefunctions obtained numerically from a tight binding Hamiltonian with those predicted by the continuum surface model, and found that while they are in agreement when no magnetic field is present, some discrepancies arise when a gauge field is applied.

Without a gauge field, i.e.\ with only a Zeeman field, the massive Dirac surface theory matches well with the lattice model. In particular, both the surface spectrum at small momentum, and the pseudospinor direction of the states at the surface band edges, show agreement between the surface theory and the lattice model. For surfaces normal to the $x$-axis, the spin orientation of the surface states, as predicted by the surface theory, is (anti)parallel to the $x$-direction, and is opposite for opposite energies and opposite surfaces.  This is supported by the lattice model and gives rise to the fact that the surface gap increases with the Zeeman field on one surface while decreasing on the other.

When a gauge field is applied, the chiral hinge modes of the SOTI persist, but can be split into two pairs in momentum space. The coupling between the applied field and the electronic orbital motion leads to the formation of Landau levels on those surfaces of a nanowire that are pierced by flux. The continuum Dirac theory successfully predicts that the lowest Landau level exists away from zero energy and in opposite directions on opposite surfaces, but overestimates the magnitude of the energy of that Landau level. 

The discrepancy means that in our lattice model, the lowest Landau level falls within the surface gap, the effect of which can be seen in transport.  When the magnetic field is along the $x$-axis, the hinge modes connect the resulting Landau levels on the surfaces normal to the $x$-axis to the massive Dirac states on the surfaces normal to the $y$-axis. A clear magnetotransport signature results: within the surface gap, depending on the energy, either one or two one-dimensional chiral channels exist in each direction. As well, restricted to one face of the nanowire, the excited Landau levels are not symmetric around zero energy.

\acknowledgements{The authors wish to thank Ashley Cook and Ranjani Seshadri for useful discussions. The work in this manuscript has been supported by NSERC and FRQNT (BAL and TPB) and by a Schulich Graduate Fellowship (BAL).}

\bibliography{hoti}
\end{document}